# Original Research Article

## Assessing prompting frameworks for enhancing literature reviews among university students using ChatGPT

Aminul Islam[1], Mukta Bansal[2], Lena Felix Stephanie[1], Poernomo Gunawan[2], Pui Tze Sian[2], Sabrina Luk[3], Eunice Tan[4], Hortense Le Ferrand[1,5*]

[1]School of Mechanical and Aerospace Engineering, Nanyang Technological University, 50 Nanyang Avenue, Singapore 639798

[2]School of Chemistry, Chemical Engineering and Biotechnology, Nanyang Technological University, 50 Nanyang Avenue, Singapore 639798

[3]School of Social Sciences, Nanyang Technological University, 50 Nanyang Avenue, Singapore 639798

[4]School of Humanities, Nanyang Technological University, 50 Nanyang Avenue, Singapore 639798

[5] School of Materials Science and Engineering, Nanyang Technological University, 50 Nanyang Avenue, Singapore 639798

[*] Hortense@ntu.edu.sg

## Abstract

Writing literature reviews is a common component of university curricula, yet it often poses challenges for students. Since generative artificial intelligence (GenAI) tools have been made publicly accessible, students have been employing them for their academic writing tasks. However, there is limited evidence of structured training on how to effectively use these GenAI tools to support students in writing literature reviews. In this study, we explore how university students use one of the most popular GenAI tools, ChatGPT, to write literature reviews and how prompting frameworks can enhance their output. To this aim, prompts and literature reviews written by a group of university students were collected before and after they had been introduced to three prompting frameworks, namely CO-STAR, POSE, and Sandwich. The results indicate that after being exposed to these prompting frameworks, the students demonstrated improved prompting behaviour, resulting in more effective prompts and higher quality literature reviews. However, it was also found that the students did not fully utilise all the elements in the prompting frameworks, and aspects such as originality, critical analysis, and depth in their reviews remain areas for improvement. The study, therefore, raises important questions about the significance of utilising prompting frameworks in their entirety to maximise the quality of outcomes, as well as the extent of prior writing experience students should have before leveraging GenAI in the process of writing literature reviews. These findings are of interest for educators considering the integration of GenAI into academic writing tasks such as literature reviews or evaluating whether to permit students to use these tools.

## Introduction

Writing a literature review is a common task for students, in both science, technology, engineering, and mathematics (STEM) and non-STEM fields. Writing a literature review requires high level cognitive ability, being a constructive process requiring conceptualization, realisation and critique (Ondrusek, 2012). Writing literature reviews can be a challenging task, as it requires students to demonstrate mastery in reading, literacy, vocabulary and synthesizing skills (Ondrusek, 2012). Reading and writing literature reviews foster the acquisition of new knowledge, identification of problems, synthesis, argumentation, interpretation and dissemination of the findings (Castillo-Martínez & Ramírez-Montoya, 2021). The task can be even more difficult when the writing is done in a non-native language, the topic is new to the student, or the student exhibits anxiety or lack of confidence. Literature reviews are essential for a successful transition from undergraduate to postgraduate studies, where research projects require the ability to examine the current state of the art in the chosen field and extract key fundamental concepts. Moreover, most postgraduate studies are evaluated through writing assignments, including academic papers and dissertations. Universities' curricula typically include literacy and writing courses, yet students often struggle with writing literature reviews. This therefore highlights the need to support students in this critical academic task.

There exists ample literature to guide students on how to write a literature review in various disciplines and at various academic levels (Baumeister, 2013; Choi et al., 2019; Walter & Stouck, 2020; Younas & Ali, 2021). These papers provide very practical tips that can help students improve their writing and deliver good literature reviews. In recent years, students have gained access to new tools that are based on artificial intelligence (AI) and Large Language Models (LLMs) which are capable of generating extensive text. These tools have been readily adopted by students for brainstorming, planning, translating and reviewing their written assignments (Levine et al., 2024). Studies have also demonstrated that the use of AI can significantly improve the quality of essays submitted by students (Guo et al., 2024; Marzuki et al., 2023; Yang et al., 2025). Notably, the most substantial improvements were observed in the structure and organisation of their papers. To effectively use the AI tools, students need to be aware of how it operates and produces the output and learn how to prompt it, which is referred to as prompt engineering (Giray, 2023; White et al., 2023). Yet, it is still not clear how students prompt the GenAI tools and how these are effective in the context of writing literature reviews.

In this study, we introduce three prompting methods to a group of students for writing a literature review using ChatGPT and then evaluate both their prompting behaviour and the quality of their literature reviews before and after exposure to the frameworks. The following specific research questions are addressed:

RQ1: How do students prompt ChatGPT, and how does awareness of specific prompting frameworks influence their prompting behaviour?

RQ2: To what extent do prompting frameworks improve the quality of students' literature reviews produced with ChatGPT, and what specific aspects show the most improvement?

After a literature review about the use of AI in education, literature review writing, and prompt engineering, the methodology for this study is described. A group of university students was tasked with writing a literature review on two research articles provided beforehand. The students then attended a two-day workshop where they were introduced three prompt engineering frameworks: CO-STAR, POSE, and Sandwich. After introducing the frameworks, the students were asked to re-write their literature reviews using ChatGPT. Overall, we found that most of the students had already used ChatGPT to help them generate the pre-workshop literature review and that their prompting behaviour changed after learning about prompting frameworks during the workshop. Significant improvements were observed in the literature reviews delivered by the students after they had re-written their reviews using the prompting frameworks introduced to them. This study provides key insights into how prompting frameworks can improve the quality of students' written assignments, while also clarifying the frameworks' potential benefits and limitations.

## Literature review

### AI in education

AI is a recent technology that is receiving a rising interest among educators for education and research. For instance, educators have leveraged AI technologies such as speech-to-text, text-to-speech, and real-time language translation to address diverse learning needs and create more inclusive educational environments (U.S. Department of Education, 2023). GenAI platforms such as Claude, Cohere, Gemini, GPT, LLaMa and Mistral, present both opportunities and challenges for education (Jauhiainen & Guerra, 2024). Notwithstanding the popularity and potential of GenAI and LLMs in academia, concerns remain about AI-assisted cheating which involves students completing their written assignments and examinations with the help of GenAI. Črček & Patekar (2023) studied a group of 201 students from various public and private universities in Croatia and reported that more than half of them used ChatGPT for their written assignments. Another study by Jurado et al. (2023) examined whether American and Swedish students in higher education differed in the extent to which they used AI tools: while over 40% of the American students studied used AI tools, the usage rate was even higher for Swedish students. Such widespread use of GenAI tools among students has brought with it growing concerns about AI-assisted cheating, an inevitable consequence, especially in the absence of proper guidance and safeguards (Hrdličková, 2024). A defining characteristic of AI-assisted cheating is students' over-reliance on GenAI, which hampers their development of critical thinking and innovation skills. This dependency could impair their judgment about the accuracy of GenAI-generated content and result in their unquestioning acceptance of GenAI outputs. As Raitskaya and Tikhonova (2025) highlight, critical thinking is key to navigating the risks posed by algorithmically generated content. Without it, students may struggle to refine prompts, interpret AI responses, or verify information accuracy. For example, students who do not actively engage in critical thinking might fail to recognize and avoid unreliable sources of information generated by AI hallucination and confabulation (Dvorchak et al., 2024), potentially compromising the quality and credibility of their academic work, particularly in research and literature reviews. AI-assisted academic dishonesty, compounded by the widespread inability of students to engage constructively with GenAI, has resulted in several educators and institutions banning GenAI tools altogether (Alasadi & Baiz, 2023; Črček & Patekar, 2023; Okaiyeto & Xiao, 2023). However, such prohibitive measures risk placing students at a disadvantage, leaving them ill-prepared for professional environments which are increasingly becoming AI-augmented (Korzynski et al., 2023). A more pedagogically sound approach would be to teach students the principles of prompt engineering (Lee & Palmer, 2025), which will enable them to discern between accurate output and hallucinations or confabulations, interrogate AI responses using precise, contextually rich inputs, and resolve any inconsistencies in GenAI outputs.

**Literature review as a critical academic skill**
A literature review is a systematic approach to identify, evaluate, and interpret work produced by researchers, scholars, and practitioners (Marshall, 2015, as cited in Baker, 2016). Literature review-based assignments are common in institutions of higher learning to get students into the practice of "critical appraisal of the current collective knowledge on a subject" (Winchester & Salji, 2016, p.308). Literature reviews necessitate that authors thoroughly educate themselves on the subject matter, which not only enhances their understanding of the subject but also helps strengthen their arguments based on current research and knowledge gaps (Denney & Tewksbury, 2013). The ability to critically appraise and synthesize ideas is a core, fundamental skill that will serve as a solid foundation for endeavours such as long-term research projects, dissertations, or independent studies (Rewhorn, 2018). The benefits of training students in writing literature reviews include solidifying knowledge in a specific topic or research area, identifying gaps in the research literature, and improving scientific writing skills (Hati & Bhattacharyya, 2024). Conducting literature reviews is an essential academic skill that involves more than just summarising published facts; it also entails developing an unbiased narrative backed by published evidence while offering one's own perspective (Winchester & Salji, 2016). A well-executed literature review demonstrates a student's proficiency in three domains: 1. Effective literature search aided by information literacy skills, 2. Language proficiency developed through exposure to diverse reading materials even beyond the research topic, improving vocabulary and grammar and 3. Critical writing that calls for analytical engagement with the material in the form of breaking down key elements, identifying key claims, evaluating relationships, and contrasting perspectives (Leite et al., 2019). Such a well-developed literature review that empowers the student to establish their own academic voice with confidence demands 'evaluation', the highest level of Bloom's cognitive skills (Granello, 2001). It is therefore imperative that students are thoroughly trained in this area to demonstrate their enduring academic abilities (Leite et al., 2019).

**AI tools for writing literature reviews**
Literature review is a rigorous, time-consuming and resource-intensive process, necessitating the development of tools to support or at least partially automate the process (Bolaños et al., 2024; Zala et al., 2024). With the arrival of AI to alleviate the burden of repetitive tasks performed by humans, researchers and scientists are keen to exploit the powerful computational techniques of AI to accelerate results (José de la Torre-López, 2023; Zala et al., 2024). Bolaños et al. (2024) examine how GenAI can partially automate systematic literature reviews (SLRs) in the screening and extraction stages. The authors conclude that there is a lot left to be desired with the current GenAI tools which are limited in their capabilities and hence not compelling enough for widespread use among researchers. They point out that these tools rely on outdated methodologies such as the use of basic classifiers which are nowhere near "state-of-the-art". In their study of the use of AI-based tools in the conduct of literature reviews, Wagner et al. (2022) note that AI-automated or -aided literature reviews require human interpretation and synthesis to facilitate insightful contributions. Zala et al. (2024) examine the integration of AI tools into SLRs within the Information Systems (IS) domain. Their study focuses on six key SLR stages: (1) problem formulation, (2) literature search, (3) screening for inclusion, (4) quality assessment, (5) data extraction and (6) data analysis and interpretation. They conclude that current AI tools are predominantly geared towards supporting paper screening and identification, with limited support for the other SLR stages. The authors envision "more nuanced and sophisticated" AI capabilities that automate the entire SLR workflow. Thus, most researchers concur that the existing AI tools cannot produce a high-quality review by themselves and are far from being fully automated (Pearson, 2024).

**Prompt Engineering Frameworks for Literature Reviews**

Prompt engineering frameworks have been developed to optimize interactions with GenAI systems by guiding the prompt design process and ensuring that outputs are contextually relevant and aligned with specific objectives. These frameworks advocate systematic approaches that emphasize clarity, relevance, and iterative refinement. Here we present three frameworks relevant to the study: CO-STAR, POSE, and Sandwich. While other frameworks exist, they tend to have a structure similar to those presented here, with prompts predominantly encompassing six major components: the goal, the type of content and format desired, the context, some more details on the desired output, about the style and some limitations or constraints that there may be (Eager et al., 2023). Each framework offers a structured set of considerations or steps, which are typically adapted to the needs of researchers, educators, and other professionals seeking reliable, high-quality AI-generated content.

*CO-STAR framework*

The CO-STAR framework was proposed by Launchpad, a Singapore government-backed organization (AI-Division, 2023; Zentennial, 2024). In this framework, the prompt is divided into six key elements: Context, Objective, Style, Tone, Audience, and Response. **Context** provides information on the circumstances that surround a given topic or issue related to the task. **Objective** specifies the primary purpose of the task. **Style** defines how the AI tool should present the content. For example, the style can be defined as educator-like, student-like, etc. **Tone** gives more details on the intended manner of delivery. For example, the tone can be academic, formal, casual, poetic, or humorous. **Audience** identifies who the intended recipients of the result are. **Response** specifies the expected parameters, such as length and format.

*POSE framework*

POSE has been developed in-house by a company MSD, Singapore, to promote clarity and precision by identifying four principal components: Persona, Output format, Style, and Example. **Persona** specifies the characteristics and behaviours that the AI model should display. This choice influences how the system addresses the query or task, whether as a mentor, an expert researcher, or a supportive peer. **Output format** includes the structural and tonal expectations for the generated text. Details such as the number of words, paragraph structure, and desired level of formality can be specified in this component. **Style** reflects the target audience or the preferred writing approach, which might be academic, explanatory, or instructive. **Example** presents concrete illustrations of the desired output, thereby guiding the AI model more effectively. For instance, when exploring methods for learning about investing, one might provide sample scenarios that highlight basic investment principles or present simplified dialogues illustrating the steps involved in sound financial planning.

*Sandwich framework*

The Sandwich framework is an iterative method where an initial draft is produced by the AI model and the resulting output is then edited or revised by the human who refines aspects such as accuracy, logical progression, and compliance with scholarly conventions. The updated draft is subsequently reintroduced into the AI system for additional feedback, suggestions, or expansions. Through multiple rounds of review and enhancement, the process converges on a final version that fulfils academic standards while benefiting from the efficiency of AI-driven text generation (Macri et al., 2023).

**Students' prompting habits**

Few studies have investigated students' prompting habits. One such study conducted by Sawalha et al. (2024) found that students tend to use three types of prompting strategies: (i) single copy and paste prompting, (ii) single reformulating prompting, and (iii) multiple question prompting. Approach (i) was the most frequently used approach, with a frequency of 47%, followed closely by approach (iii) at 40%, while approach (ii) was the least used at 13%. The students using approach (iii) achieved better results in their assignments. In another study by Tassoti (2024), 60% of students used approach (i), 30% used approach (ii), and 10% adopted approach (iii). Here again, approach (iii) was perceived to produce higher-quality AI outputs. Focussing on prompts used for Python coding, Denny et al. (2024) developed an exercise called 'Prompt problems', where students had to develop prompts to solve coding problems. Some students found that crafting effective prompts helped them strengthen their computational thinking skills, although there was also some reticence in putting effort into the prompts. A more recent study by Apriani et al. (2025) demonstrated that ChatGPT significantly enhanced the academic writing capabilities of undergraduate students in an Indonesian university. Across 14 guided sessions, students were introduced to ChatGPT's functionalities and instructed to use it for drafting a condensed undergraduate thesis. However, no structured prompting frameworks were applied or tested. Overall, the studies show that writing prompts is an emerging competency for students and that strong prompt engineering skills markedly improve the quality of AI-generated output (Tassoti, 2024; Wang et al., 2024).

**Research gaps and motivations for the current study**

Despite the significant concerns regarding AI use by students such as data privacy, algorithmic biases, diminished critical thinking, and academic integrity (Bahroun et al., 2023), a pragmatic approach in today's technological landscape, will be to embrace AI technologies such as GenAI and LLM, and teach students how to use them effectively and ethically instead of prohibiting their use (Alasadi & Baiz, 2023). As Mali (2025) notes in a systematic literature review of EFL/ESL (English as a Foreign Language and English as a Second Language) learners using ChatGPT for writing tasks, proper guidance in using the GenAI tool can enhance writing proficiency, boost motivation, and promote learner autonomy, underscoring the need for thoughtful pedagogical integration. Given that these transformative technologies are too significant to evade in the real world, the goal should be to successfully integrate them in curriculum design (Jauhiainen & Guerra, 2024). Prompt engineering plays a pivotal role in unleashing the power of LLMs (Chen et al., 2023), particularly in supporting text-based academic tasks such as literature reviews. It involves providing carefully crafted instructions to LLMs (Sahoo et al., 2024) to guide their responses to be more accurate, relevant, and context-sensitive (Chen et al., 2023). However, designing effective prompts to guide or constrain the model's generative behaviour remains a challenge (Korzynski et al., 2023). As highlighted by Li (2023), the LLMs bring with them legal and ethical risks stemming from stochastic parrots and hallucination. While 'stochastic parrots' refers to the LLM's propensity to reproduce training data without a deeper understanding, 'hallucination' refers to the LLMs generating fabricated information that deviates from reality (Li, 2023). In light of these risks, prompt engineering is an essential digital competency (Korzynski et al., 2023) that can equip students to formulate precise and purposeful prompts that yield reliable, focused, and high-quality results in tasks such as literature reviews. It is therefore important to introduce to students structured approaches like prompt engineering frameworks to strengthen their AI literacy and mitigate the aforementioned risks. Equally important is examining the effectiveness of these frameworks in producing high-quality literature reviews. However, due to the nascent nature of this field, there is a paucity of studies in these areas, highlighting a significant motivation for the present study.

## Materials and Methods

**Overall approach**
The approach taken to address the research questions is as follows. First, students were tasked with writing a literature review (LR1); their prompts (if any) and the quality of the submitted reviews were then analysed. Next, the students were taught prompt engineering frameworks and instructed to re-write the literature review entirely using GenAI, generating a second version (LR2). Their prompts and the quality of their revised reviews were once again analysed. For their first literature review (LR1), the students could choose whether to use ChatGPT but had to declare their choice. For their second literature review (LR2), however, using ChatGPT was mandatory.

Two research papers were selected for the literature review: Kwak et al. (2022) and Lee et al. (2020), which are related to the topic "Health institution support for the elderly". These two papers were carefully selected for their accessibility to participants from diverse backgrounds. Additionally, the papers are concise and relevant, with a focus adapted to the context of Singapore, where the study is being conducted. Instructions for conducting the reviews included following a clear structure comprising an introduction, a body with subheadings, a conclusion, and a bibliography. The participants were given one hour to write a literature review of 500 to 800 words based on the two journal articles. The participants were also made aware of the assessment rubric that would be used to evaluate the literature reviews. This rubric included criteria such as comprehensiveness and depth of the review, critical analysis, clarity and organization, writing style, originality, and significance, and is described in detail in the subsequent sections.

**Participant demographics**
A group of 7 students participated in the study. Despite the small sample size, the study yielded a rich dataset comprising the two literature reviews and all associated prompts. The participants' demographics were collected via an online survey prior to the research activity. The participants were a mix of undergraduate (third or final year), master's, and PhD students from Nanyang

Technological University, Singapore (**Figure 1**). Most participants had a STEM background, with over half of them being master's students (**Figure 1a,b**). Most participants were between the age range 18 and 30 years old, with about one third older than 30 but below 60 (**Figure 1c**). Regarding their familiarity with GenAI tools, most of the participants reported using such tools somewhat frequently, while the remaining participants indicated that they always use them (**Figure 1d**). The students indicated various levels of familiarity with writing literature reviews (**Figure 1e**). On a scale of 5, the students on average reported a high level of difficulty with the various tasks involved in literature reviews (**Figure 1f**).

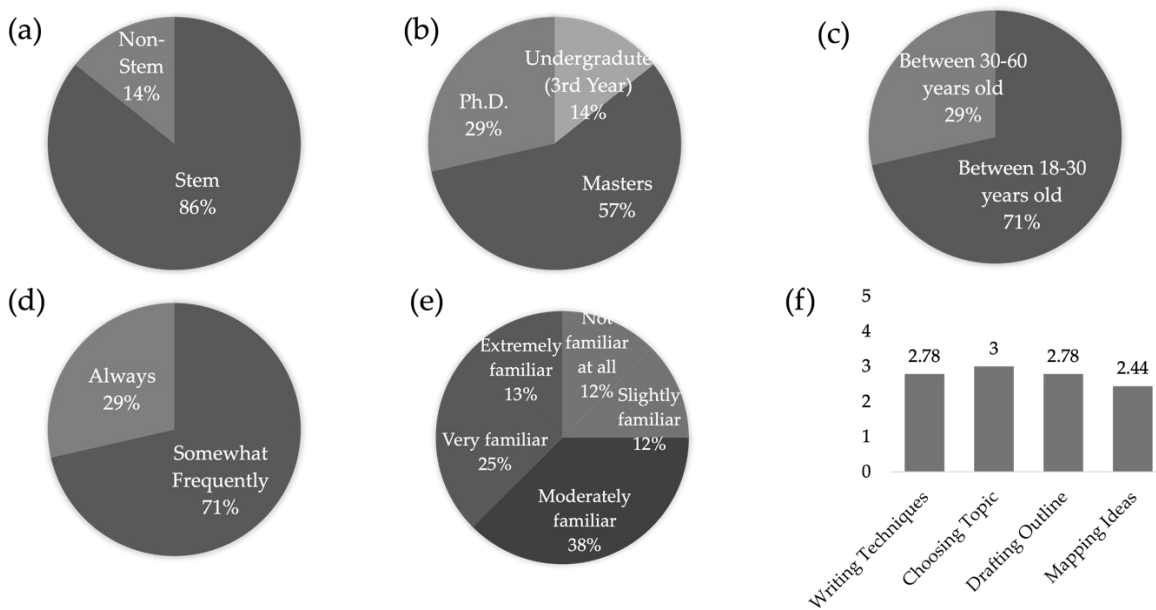

**Figure 1: Participants demographics. (a)** Area of study, **(b)** level of study, **(c)** age, **(d)** frequency of usage of GenAI tools for academic writing, **(e)** familiarity with writing literature reviews, **(f)** mean level of difficulty reported by the participants across various elements of literature review writing.

**Analysis of the prompts**

The prompts used by the participants were collected using the software Genue.AI (Genue, Singapore) in which ChatGPT0.4 is embedded and allows automatic recording of prompts and answers. They were analysed using the rubrics shown in **Table 1**, **Table 2** and **Table 3**, for the frameworks CO-STAR, POSE, and Sandwich, respectively. The number of prompts and the number of words used in the prompts were also counted. Citations included in the prompts were excluded from the word count.

**Table 1:** CO-STAR framework evaluation rubric

| Elements | Performance indicator | | |
|---|---|---|---|
| | 0 (Below Average) | 1 (Average) | 2 (Above Average) |
| Context (C) | Lacks or omits essential background. Too vague for the AI to grasp the setting. | Minimal background provided. Some relevant information but may still be unclear or incomplete. | Thorough, explicit background. Clearly situates the topic with key details (who, what, when, where, why). |
| Objective (O) | No clear goal/ objective stated. Leaves the AI to analyze the objective/goal. | Basic mention of a goal but lacks clarity or specifics. The AI has some ideas but may need more detail. | Crisp, well-defined objective directly tied to the context. The AI can analyze exactly what is expected. |
| Style (S) | No role or persona specified. Defines wrong/incorrect roles | Vaguely mentions a role but does not clarify domain or level of expertise. | Clear, relevant persona. Aligns with context and objective, making it easier to judge or trust the response. |
| Tone (T) | No tone specified. Possibly conflicting or confusing tone cues. | Basic indication of tone but not fully integrated. | Explicitly states and maintains tone. |
| Audience (A) | No mention of the audience. Used prompt that might make AI produce generic or mismatched content. | Audience hinted at, but not clearly defined. AI can partially adapt but lacks full clarity on audience needs. | Clearly identifies the audience's background or needs. AI can fully customize the response. |
| Response (R) | No guidance on format or length. AI output may be disorganized or not aligned to user's purpose. | Some mention of format but not much detail on structure or length. | Detailed instructions on how to format. Given precise instructions for AI to structure and present the final text. |

**Table 2**: POSE framework evaluation rubric

| **Elements** | **Performance indicator** | | |
|---|---|---|---|
| | **0 (Below Average)** | **1 (Average)** | **2 (Above Average)** |
| Persona (P) | No clear persona or role. Missing traits or behavioral cues. The prompt used would force AI to guess its viewpoint, leading to generic or mismatched responses. | Persona is named, but vague or incomplete. Some traits or behaviors are mentioned but lack full alignment with the task. Gives AI a semi-targeted way to operate | Persona is explicitly defined. Traits and behaviors clearly match the intended task/context. Enables AI to fully immerse in the role and deliver |

10
PROMPT ENGINEERING TO IMPROVE LITERATURE REVIEWS| | | but may need more direction. | context-specific responses. |
|---|---|---|---|
| Output Format (O) | No clear instructions on structure, tone, or length. The AI's response may be disorganized, too brief, or too long. | Some mention of structure or tone but lacks thorough details. | Detailed instructions on format. Specific tone request. Clear word count range or approximate length, ensuring the AI produces the desired scope and style. |
| Style (S) | No audience specified. The language could be too generic or mismatched. | Audience is hinted at but not in detail. The AI can partially adapt but may not fully meet audience expectations. | Clear audience definition. Language, examples, and complexity are tailored precisely to the audience's level. |
| Example (E) | Does not provide examples. The AI may produce abstract or theoretical text without real-world illustrations. | Prompts for at least one example, but not clearly detailed. The examples might be generic or limited in depth. | Asks specifically for domain-relevant examples. Offers relevant information or parameters for the application examples. |

**Table 3**: Sandwich framework evaluation rubric

| **Elements** | **Performance indicator** | | |
|---|---|---|---|
| | 0 (Below Average) | 1 (Average) | 2 (Above Average) |
| Initial Prompt | Vague or minimal prompt. | Basic prompt with limited clarity. | Clear, targeted prompt for initial response. |
| Prompt for AI feedback | No real review requested. No revision based on feedback. | Generic request for review. Did not ask AI for suggestions. | Detailed request for expert critique. Meaningful revisions based on AI's constructive feedback. |

**Analysis of the literature reviews**
The literature reviews LR1 and LR2 produced by the students were collected. iThenticate was used to generate similarity reports for LR1 and LR2 with the existing literature as well as similarity between LR1 and LR2. All LR1 and LR2 reviews were also assessed according to the rubric presented in **Table 4.** Three independent examiners evaluated all 14 literature reviews using the rubric. For the quantitative analysis, the average scores of all major and sub-criteria were calculated separately for students who used GenAI in LR1 and for those who did not use GenAI. The examiners also provided overall evaluative feedback on the quality of the literature reviews.

**Table 4**: Literature review evaluation rubric

| Criteria | Performance indicator | | |
| --- | --- | --- | --- |
| | **Below average: 0** | **Average: 1** | **Above average: 2** |
| **Content** | | | |
| Comprehensiveness and depth of the literature review | Does not clearly define the topic of the review. Subtopics are not relevant to the review main topic. | Clearly defines review topic and subtopics. Some major themes are missing. | Demonstrates a thorough understanding of the relevant literature on the topic. Identifies and discusses key themes, theories, and debates within the field. |
| Accuracy of the literature content | The content shows inaccuracies with false statements. | The content is globally accurate, but some statements are ambiguous. | The content is accurate without any false statements. |
| Critical analysis | Makes no effort to find connections between studies. Fails to identify strengths and weaknesses. | Makes some connections between studies with basic analysis of strengths and weaknesses. | Offers insightful connections between studies. Provides a reasonable analysis of strengths and weaknesses. Identifies gaps in the literature and suggests directions for future research. |
| **Methodology** | | | |
| Search strategy and selection of sources | Fails to mention the sources throughout review. | Sources are mentioned and referred to in the review but not always accurately. | The sources are mentioned appropriately when it is relevant. |
| **Communication and Presentation** | | | |
| Clarity | The paragraphs are not logically connected, and some sentences do not convey clear arguments. | The ideas lack connection and smooth flow which does not always convey the right meaning. The reader has some doubts about what the global idea is. | Uses logical transitions to connect ideas and arguments, making the review easy to read and to understand. |
| Organization | The review is not clearly organized with introduction, conclusion, and subheadings. | The organization of the review is missing some subheadings, or some paragraphs are misplaced. | Presents a clear and well-organized literature review with introduction, conclusion and subheadings |
| Writing style | The review is not written in an academic writing style. | The writing style of the review is not always to the academic standard. | Uses appropriate academic writing style and conventions. Maintains a consistent and professional tone throughout the review. |
| Vocabulary | The vocabulary is very general and not specialized. | Uses some technical vocabulary but terms are not. explained or are misused. | Uses specific vocabulary and explains the difficult terms to a lay audience. |
| **Overall impression** | | | |
| Originality and Insightfulness | The topics discussed are not original and do not provide a new view on the topic. | The review provides some new aspects to the topic but does not raise critical questions or challenges. | Demonstrates a fresh and insightful perspective on the topic. Raises new questions or challenges existing assumptions. Makes a significant contribution to the understanding of the topic. |
| Significance and applicability | The literature review is not connected to real-world applications and does not conclude with concrete implications or recommendations for future research or practice. | The review is poorly connected to the real-world applications and some future research recommendations are not entirely relevant. | Connects the literature review to real-world issues or applications. Provides practical implications or recommendations for future research or practice. Demonstrates the relevance and significance of |

literature to the broader field.

## Results

**Students' prompting behaviour**

To better understand how students use the GenAI tools to obtain literature reviews, we first analysed the prompts they use to obtain their literature reviews, before (**Figure 2**) and after (**Figure 3**) being introduced to the three prompting frameworks (CO-STAR, POSE and Sandwich). For both literature reviews, we analysed the prompts according to the frameworks although the participants had not been introduced to them for LR1. The motivation for doing so stems from the difficulty of evaluating a specific prompting strategy in the absence of a framework. Besides, students naturally use some components of these frameworks already as they are intuitive, which allows us to observe how key elements in the frameworks are prompted differently between LR1 and LR2.

For the five participants who used GenAI to help them generate LR1, there was little evidence to suggest familiarity with the three prompting frameworks. Examples of prompts are: "please give me some subtheme about "Health Institutions' Support to the Elderly" in Singapore" or "Write a literature review for a research topic about Health institution support to the elderly. The literature review is expected to have: (1) an introduction where the topic is introduced and the organisation of the review is described, (2) a body that should include subheadings with at least 2 subtopics, (3) a conclusion that summarizes the results and provides a perspective highlighting gaps and challenges, (4) a Bibliography". Using the rubrics corresponding to these frameworks (Tables 2, 3, and 4) to analyse the prompts, it was found that the students tended to use only 'Context' and 'Objective' elements of the CO-STAR framework (**Figure 2a**), the 'Output' element of the POSE framework (**Figure 2b**), and primarily an initial direct prompt in the case of the Sandwich framework (**Figure 2c**). This indicates that the participants' use of GenAI mostly consisted of directly asking it to provide the output, without giving much detail or context and without seeking feedback. However, the participants did use multiple prompts to refine their answers, ranging from 5 to 15 prompts, with each prompt being relatively long, typically about 50 to 300 words (**Figure 2d**).

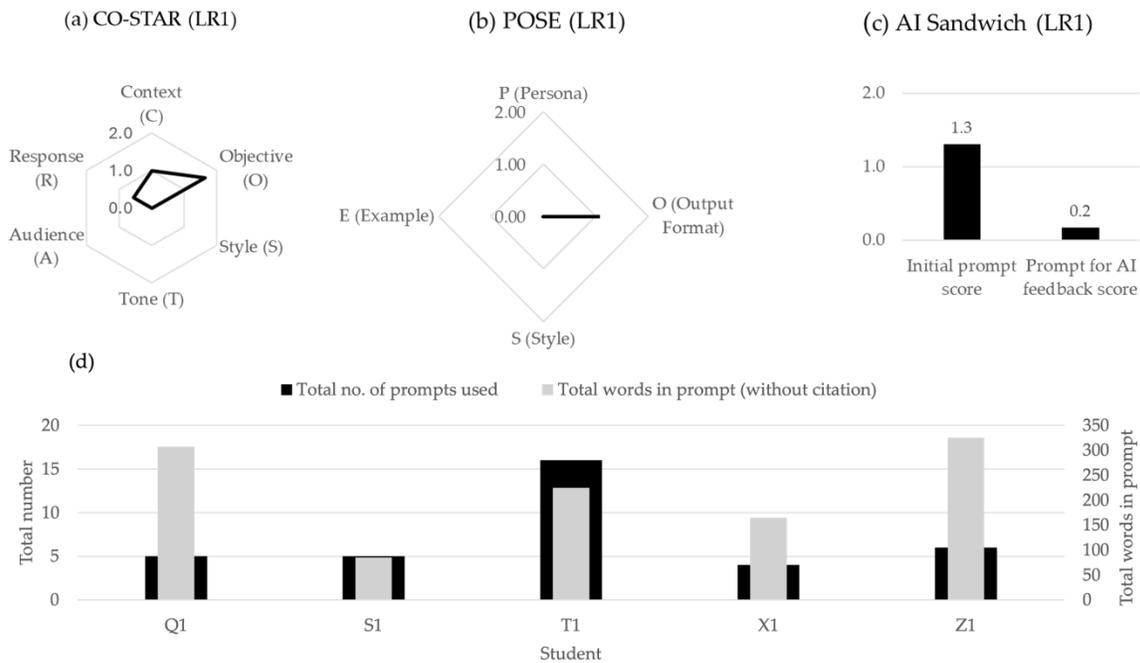

**Figure 2: Analysis of the prompts used for LR1.** Average marks for each element from the rubric tables for **(a)** CO-STAR, **(b)** POSE, and **(c)** Sandwich frameworks, and **(d)** total number of prompts and total words used in the prompts for each participant (Q1, S1, etc. represent the individual students who used ChatGPT in LR1).

Analysing the prompts used for writing LR2 (**Figure 3**), it was again not always possible to determine exactly which specific framework the participants used, as they often seemed to use a mix of elements from more than one framework. Examples of prompts are: "I am doing a student assignment of writing a literature review of about 800 words, I want a literature review for the topic: 'Health institution support to elderly'. The tone should be similar to that of a scientific paper and it will be submitted to the course instructor it should follow the following structure: the l The literature review is expected to have: (1) an introduction where the topic is introduced and the organisation of the review is described, (2) a body that should include subheadings with at least 2 subtopics, (3) a conclusion that summarizes the results and provides a perspective highlighting gaps and challenges, (4) a Bibliography."; "I have a draft of a literature review. please edit it with the following points: Context: academic literature review; Objective: giving readers an overview of the challenges to collaborative care. tone: formal; response format: detailed". In particular, the Sandwich framework was always used in parallel with the other frameworks. Overall, we found that all participants used the CO-STAR prompt framework at some point in their series of prompts. However, not all elements of the framework were used. 'Context' and 'Objective' were the most commonly included, while the remaining elements were often absent (**Figure 3a**). Similarly, among the 2 participants who used the POSE framework, 'Output Format' and 'Style' were included, but 'Persona' and 'Examples' were rarely present (**Figure 3b**). The Sandwich method was also used alongside other frameworks, with none of the participants using it exclusively (**Figure 3c**). The prompts often lacked clarity. Furthermore, we observed that the participants used fewer than six prompts, with each prompt averaging around 100 words (**Figure 3d**).

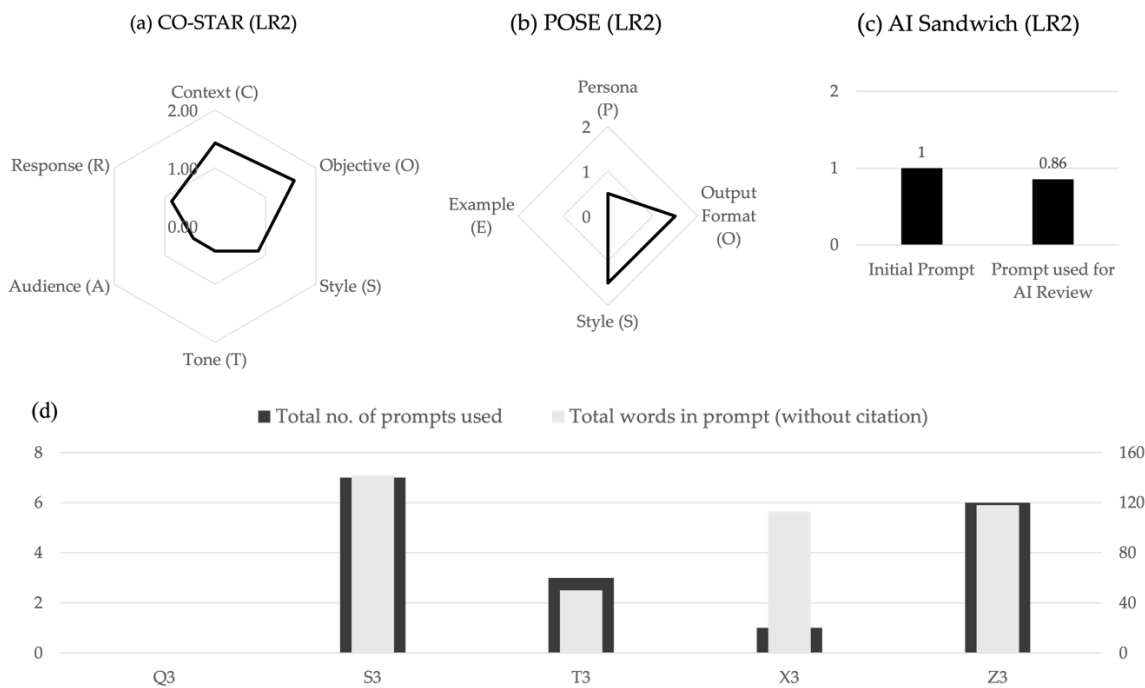

**Figure 3: Analysis of the prompts used for LR2.** Average marks for each element from the rubric tables for **(a)** CO-STAR, **(b)** POSE, and **(c)** Sandwich frameworks, and **(d)** total number of prompts and total words used in the prompts for each participant (Q3, S3, etc. represent individual students).

This analysis highlights a slight shift in the prompting behaviour of the participants before and after being introduced to the prompting frameworks. After exposure to these frameworks the participants used fewer prompts and incorporated more elements from the frameworks into their prompts. This suggests that the simplicity and structure offered by the frameworks may have been appealing to the students. Indeed, the frameworks provide a very clear structure which could readily be used as a template by the participants. The participants used the CO-STAR framework more frequently, which was consistent with the post-workshop feedback we received from them. Despite this, we noted that the frameworks were not used in their entirety or always applied appropriately. This could be due to several reasons: (i) the results generated were satisfactory to the participants, so they did not feel the need to add more details, especially related to the tone or persona elements; (ii) the frameworks may not have been fully understood by the participants. These observations raise questions about the robustness and suitability of the frameworks for writing literature reviews, as well as the students' ability to critically analyse their own work.

Next, we analyse the literature reviews that were the output from these prompts in order to explore whether this shift in prompting behaviour had an impact on the outcomes.

**Literature review outputs**
The aim of this study is to help students produce high quality literature reviews. Accordingly, the literature review outputs from LR1 and LR2 were assessed (**Figure 4**) using the rubric previously defined in **Table 4**.

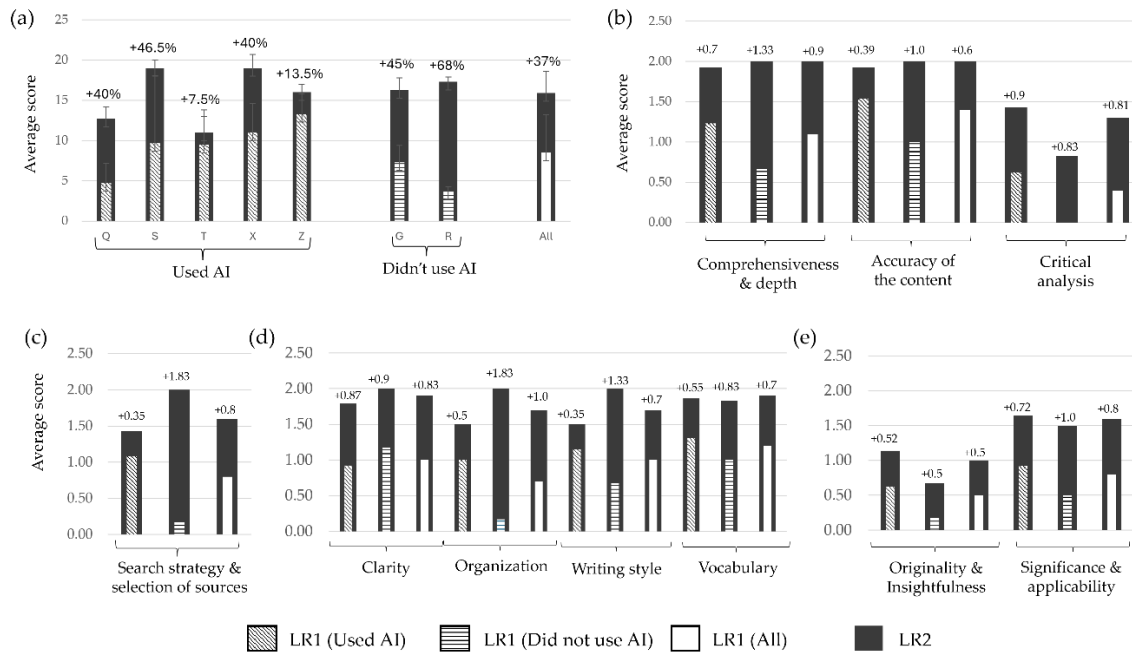

**Figure 4: Analysis of the literature reviews LR1 and LR2. (a)** Total score for each participant (Q, S, etc. represent individual students), **(b)** average score for content, **(c)** average score for methodology, **(d)** average score for communication and presentation, **(e)** average score for overall impression. LR1 from participants who did not use AI, pattern with crossed lines; LR1 from participants who used AI, pattern with horizontal lines; LR1 overall, white; LR2 overall, black.

Comparing the total score for each participant in LR1 and LR2, an overall improvement of approximately 37% was observed in their marks for LR2 compared to LR1, with even greater improvement in the literature reviews produced by students who did not use any GenAI for their LR1 (**Figure 4a**). The greater improvement observed in the literature reviews produced by participants who did not use GenAI in their LR1 is also evident across each specific category of the evaluation rubric. The only exception was clarity, which was scored slightly higher for students who did not use GenAI in LR1 as compared to those using GenAI in their LR1. However, due to the sample size it is not possible to determine if those students not using GenAI had more writing proficiency or if this is just a chance occurrence. Across all students, comprehensiveness, depth, and accuracy increased in LR2 (**Figure 4b**). However, critical analysis remained weak with an average score below 1.5 for all reviews. One hypothesis for this result could be the lack of knowledge of the students in the specific area of the topic of the reviews, or lack of time for critical thinking. Structuring and articulating methodological approaches also improved in LR2 (**Figure 4c**). Similarly, clarity, organisation, writing style, and vocabulary increased in LR2 (**Figure 4d**), as well as originality and insightfulness, and significance and applicability (**Figure 4e**).

While LR2 showed significant improvement over LR1 based on the evaluation rubric, additional comments were also collected from the evaluators for each participant (**Table 5**). This qualitative analysis of the reviews further highlighted improvements in structure, organisation, and clarity in LR2 compared to LR1. Some participants included diagrams and frameworks in their LR2, and some reviews addressed future research directions or policy implications. However, several issues

17
PROMPT ENGINEERING TO IMPROVE LITERATURE REVIEWSremained, such as missing or incorrect references and citations, or over-reliance on a single source. Moreover, most LR2 exceeded the word limit while the rest had inconsistent formatting. While the polished and formulaic style of AI-generated text made the reviews more coherent and readable, it often lacked deeper critical analysis, originality, or engagement with the strengths and weaknesses of the material. Most LR2 submissions appeared to be predominantly generated using GenAI. This was particularly visible in the structure and organisation of the literature reviews, with extensive use of subtitles and sub-headings as well as terms like "body".

**Table 5:** Comments on the literature reviews from participants who used ChatGPT in LR1 and LR2.

| Participant | Comments on their literature reviews' structure and content | |
|---|---|---|
| | **LR1** | **LR2** |
| G | <ul><li>Basic summarization</li><li>No reference</li><li>No title sections</li><li>Only 427 words</li></ul> | <ul><li>More structured, detailed and comprehensive.</li><li>Relies on one paper only (Lee et al. (2020)).</li><li>Has a table of contents</li><li>More than 800 words</li><li>4 references</li></ul> |
| R | <ul><li>Poorly organized</li><li>No reference.</li><li>Formatting issues (presence of asterisks).</li><li>Only 481 words</li></ul> | <ul><li>More than 800 words</li><li>2 references</li></ul> |
| Q | <ul><li>Reads like an outline than a thorough review</li><li>Lacks depth.</li><li>Only 427 words</li><li>No reference</li></ul> | <ul><li>Relies heavily on one paper (Lee et al. (2020)).</li><li>Includes a diagram</li><li>More than 800 words</li></ul> |
| S | <ul><li>Only reviewed the first paper (Kwak et al. (2022))</li><li>Is incomplete.</li><li>Only 467 words</li></ul> | <ul><li>More cohesive, connects the papers.</li><li>No new insights.</li><li>More than 800 words</li><li>3 references</li></ul> |
| T | <ul><li>Citation mistake</li></ul> | <ul><li>Only slightly improved from LR1, but it is largely similar in content.</li></ul> |
| X | <ul><li>Only 567 words</li></ul> | <ul><li>More than 800 words</li></ul> |
| Z | • Have missing citation and used wrong reference.<br>• Only 617 words | • Have missing citation and used wrong reference. |

## Discussion

**Correlation between the prompting framework and the literature review scores**

This study aims to understand how the prompting behaviour of students and their use of prompting frameworks can help them produce high quality literature reviews. We therefore present plots to compare their prompting and literature review scores from LR1 and LR2 **(Figure 5).** The two graphs represent the CO-STAR and POSE frameworks, with scores averaged across participants who used these frameworks for LR2 and compared against their LR1 scores. The Sandwich framework was not used to study the correlation as no participant used this framework exclusively. Overall, we observed that between LR1 and LR2, the average prompt scores increased from 19% to 46% for the CO-STAR framework and from 11% to 46% for the POSE framework. Correspondingly, the literature review scores increased from 43% to 80% for CO-STAR and from 61% to 88% for POSE. The use of these prompting frameworks, despite being incomplete, still resulted in a significant increase in the grades obtained by participants for their literature reviews, demonstrating their positive impact. The extent of improvement was even greater among participants who used the CO-STAR framework, as they had lower LR1 scores as compared to those who used the POSE framework. However, their LR2 scores were comparable, ranging between 80% and 88%. This suggests that either framework may be effective in helping students improve their literature reviews.

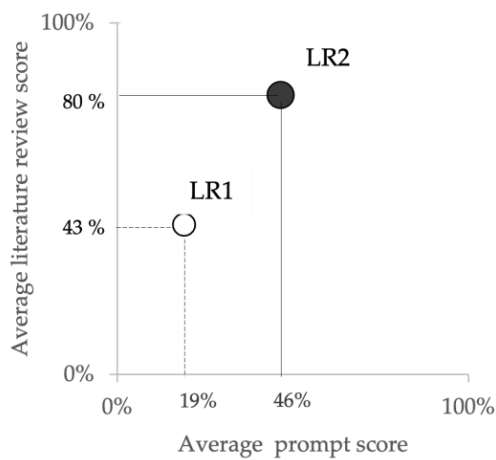 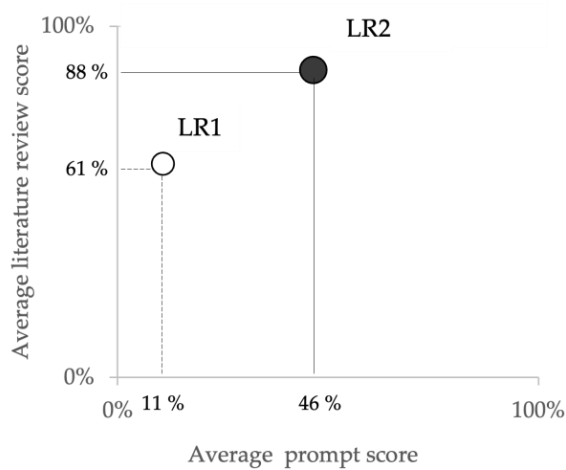

**Figure 5:** Correlation between literature review and prompt scores for LR1 and LR2 using **(a)** the CO-STAR framework, and **(b)** the POSE framework.

**Similarity of the literature reviews**

While an improvement in literature review scores was observed for all students after using the prompting frameworks (LR2), the extent to which this improvement depends on the quality of the initial draft (LR1) remains to be determined. Also, whether the originality of the literature review in LR1 was in any way altered in LR2 needs to be assessed. Given that GenAI tools like ChatGPT rely on publicly available data, there is a risk of unintentional plagiarism. The absolute similarity percentages of LR1 and LR2 were assessed, along with the similarity of each LR2 relative to its corresponding LR1, using iThenticate (**Figure 6**). It was observed that the percentage of similarity with existing literature was higher for LR2 compared to LR1 for all reviews, ranging from 10% to 20 % (**Figure 6a**). Universities typically set the acceptable similarity threshold at around 15%, suggesting that the use of AI tools may risk exceeding the ethical limits on plagiarism. Furthermore, LR2 showed a high degree of similarity to LR1 especially when LR1 had already been generated using AI (**Figure 6b**). This suggests that LR2 was, in effect, another iteration of LR1, albeit refined using the prompting frameworks. So, although the students were expected to start their LR2 'from zero', it appears that, in reality, they either reused content from their LR1, or the model retained memory of it. Finally, the similarity level when comparing the individual LR2 together and with each other is only about 6% (**Figure 6c**). This suggests that despite the evaluators' impressions of similarity across all LR2 submissions, each literature review was in fact distinct. This may suggest that the tone of the generated writings and the style were similar, but not the same words were used.

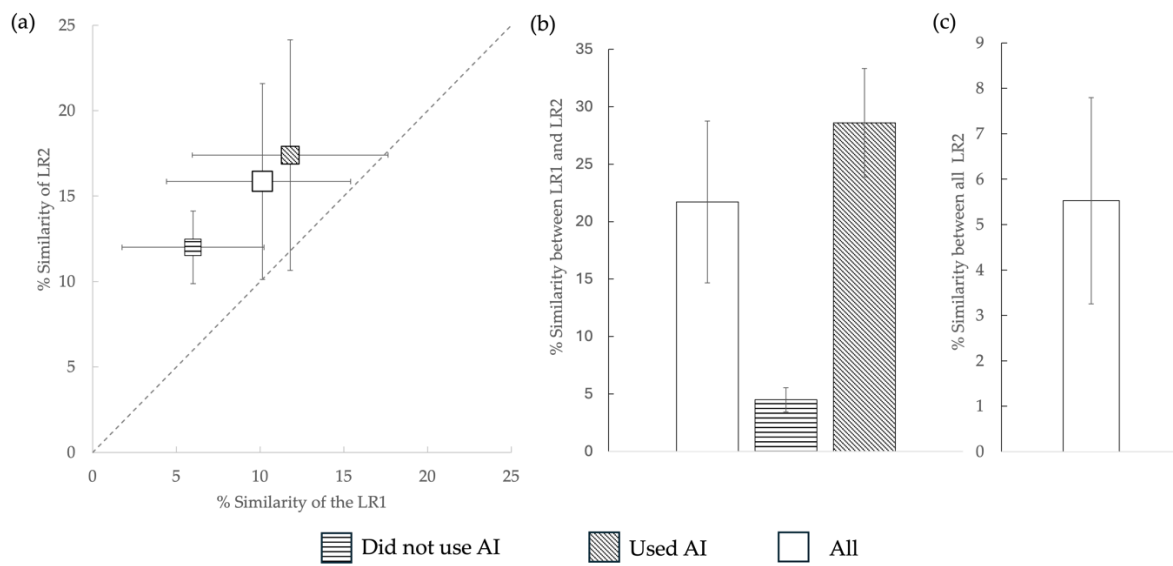

**Figure 6: Similarity analysis of LR1 and LR2 submissions. (a)** Comparison of similarity percentages of LR1 and LR2 with existing literature. **(b)** Similarity between LR1 and LR2. Participants who did not use AI, horizontal lines pattern; participants who used AI, tilted lines pattern; all students, white. **(c)** Similarity between all LR2.

## Conclusion

This study investigated the prompting behaviour of students when writing literature reviews and how the use of prompting frameworks helped them improve their reviews. The data were collected from university students at various academic levels. Prompts and literature reviews produced by the participants before and after being introduced to three prompting frameworks using ChatGPT were compared. The results showed that while participants did not fully adopt the introduced prompting frameworks when generating their second literature reviews (LR2), the outcomes still demonstrated significant improvement over the first round of literature reviews (LR1). However, it was noted that critical analysis, insightfulness and overall depth of discussion were areas that could be further strengthened. The improvement in the quality of the literature reviews in the second round (LR2) showed a positive correlation with the use of more effective prompts. However, the similarity percentage with existing sources also increased in the second round (LR2). These findings raise questions about the efficacy of prompting frameworks and suggest that a clearer understanding of participants' expectations and their proficiency in conducting literature reviews may be needed for better evaluation of these frameworks. This study can help educators reflect on how to better support students in writing literature reviews and in evaluating literature reviews generated with the help of GenAI.

## Acknowledgments

We acknowledge Raymond Chan Wai Mun and Angelina Teo Kwee Gaik from Genue, Singapore, for providing access to the software GenueAI. We acknowledge Jason Tamara Widjaja from MSD, Singapore, for providing the POSE framework. We acknowledge Nanyang Technological University (NTU, Singapore) for funding this project under the EdeX Faculty Learning Communities Grant. The data were collected and stored under the IRB-2023-1041.

## Conflict of interests

The authors declare that they have no conflict of interest.

## Authors contributions

A.I.: Data curation, Formal analysis, Visualisation, Investigation, Writing- original draft.

M.B.: Data curation, Formal analysis, Investigation, Methodology, Writing-original draft, Writing-review and editing.

L.F.S.: Methodology, Writing-original draft, Writing-review and editing.

P.G.: Methodology, Investigation, Writing-review and editing.

P.T.S.: Methodology, Writing-original draft, Writing-review and editing.

S.L.: Methodology, Writing-original draft, Writing-review and editing.

E.T.: Conceptualization, Funding acquisition, Project administration, Methodology, Writing-original draft, Writing-review and editing.

H.L.F.: Conceptualization, Formal analysis, Supervision, Methodology, Writing-original draft, Writing-review and editing.

**Data availability**

Data available upon request to the authors.

**References**

AI-Division, G. D. S. (2023). *Prompt Engineering Playbook*. https://www.developer.tech.gov.sg/products/collections/data-science-and-artificial-intelligence/playbooks/prompt-engineering-playbook-beta-v3.pdf

Alasadi, E. A., & Baiz, C. R. (2023). Generative AI in Education and Research: Opportunities, Concerns, and Solutions. *Journal of Chemical Education*, *100*(8), 2965-2971. https://doi.org/10.1021/acs.jchemed.3c00323

Apriani, E., Daulay, S.H., Aprilia, F., Marzuki, A.G., Warsah, I., Supardan, D., & Muthmainnah. (2025). A mixed-method study on the effectiveness of using ChatGPT in academic writing and students' perceived experiences. *Journal of Language & Education*, 11(1),26-39. https://doi.org/10.17323/jle.2025.17913

Bahroun, Z., Anane, C., Ahmed, V., & Zacca, A. (2023). Transforming Education: A Comprehensive Review of Generative Artificial Intelligence in Educational Settings through Bibliometric and Content Analysis. *Sustainability*, *15*(17), 12983. https://www.mdpi.com/2071-1050/15/17/12983

Baker, J. D. (2016). The Purpose, Process, and Methods of Writing a Literature Review. *AORN Journal*, *103*(3), 265-269. https://doi.org/10.1016/j.aorn.2016.01.016

Baumeister, R. F. (2013). Writing a Literature Review. In M. J. Prinstein (Ed.), *The Portable Mentor: Expert Guide to a Successful Career in Psychology* (pp. 119-132). Springer New York. https://doi.org/10.1007/978-1-4614-3994-3_8

Bolaños, F., Salatino, A., Osborne, F., & Motta, E. (2024). Artificial intelligence for literature reviews: opportunities and challenges. *Artificial Intelligence Review*, *57*(10), 259. https://doi.org/10.1007/s10462-024-10902-3

Castillo-Martínez, I. M., & Ramírez-Montoya, M. S. (2021). Research Competencies to Develop Academic Reading and Writing: A Systematic Literature Review [Systematic Review]. *Frontiers in Education*, *5*. https://doi.org/10.3389/feduc.2020.576961

Chen, B., Zhang, Z., Langrené, N., & Zhu, S. (2023). *Unleashing the potential of prompt engineering in Large Language Models: a comprehensive review*.

Choi, A. R., Cheng, D. L., & Greenberg, P. B. (2019). Twelve tips for medical students to conduct a systematic review. *Medical Teacher*, *41*(4), 471-475. https://doi.org/10.1080/0142159X.2018.1426847

Črček, N., & Patekar, J. (2023). Writing with AI: University Students' Use of ChatGPT. *Journal of Language and Education*, *9*, 128-138. https://doi.org/10.17323/jle.2023.17379

Denney, A. S., & Tewksbury, R. (2013). How to Write a Literature Review. *Journal of Criminal Justice Education*, *24*(2), 218-234. https://doi.org/10.1080/10511253.2012.730617

Dvorchak, J., Mistry, S. (2024). *Artificial Intelligence in education friend or foe?* Communicating Cambridge's Approcah to Generative AI, UK. https://www.cambridgeinternational.org/Images/710406-artificial-intelligence-in-education-friend-or-foe-sanjay-mistry-jesse-dvorchak.pdf

Eager, B., Brunton, R. (2023). Prompting Higher Education Towards AI-Augmented Teaching and Learning Practice. *Journal of University Teaching and Learning Practice*, *20*(5). https://doi.org/10.53761/1.20.5.02

Garrote Jurado, R., Pettersson, T., & Zwierewicz, M. (2023, 2023). *Students' attitudes to the use of artificial intelligence* ICERI2023 16th International conference of Education, Research and Innovation, Seville, Spain, November 13-15, 2023., Seville, Spain. http://urn.kb.se/resolve?urn=urn:nbn:se:hb:diva-30938

Giray, L. (2023). Prompt Engineering with ChatGPT: A Guide for Academic Writers. *Annals of Biomedical Engineering*, *51*(12), 2629-2633. https://doi.org/10.1007/s10439-023-03272-4

Granello, D. H. (2001). Promoting cognitive complexity in graduate written work: Using Bloom's taxonomy as a pedagogical tool to improve literature reviews. *Counselor Education and Supervision*, *40*(4), 292-307. https://doi.org/10.1002/j.1556-6978.2001.tb01261.x

Guo, K., Pan, M., Li, Y., & Lai, C. (2024). Effects of an AI-supported approach to peer feedback on university EFL students' feedback quality and writing ability. *The Internet and Higher Education*, *63*, 100962. https://doi.org/10.1016/j.iheduc.2024.100962

Hati, S., & Bhattacharyya, S. (2024). Writing a literature review as a class project in an upper-level undergraduate biochemistry course. *Biochem Mol Biol Educ*, *52*(3), 311-316. https://doi.org/10.1002/bmb.21814

Hrdličková, Z. (2024). *Identifying and responding to Artificial Intelligence in evaluating written assignments*. https://doi.org/10.4995/EuroCALL2024.2024.19104

Jauhiainen, J. S., & and Garagorry Guerra, A. Generative AI in education: ChatGPT-4 in evaluating students' written responses. *Innovations in Education and Teaching International*,62 (4):1377-1394. https://doi.org/10.1080/14703297.2024.2422337

José de la Torre-López, A. R., José Raúl Romero. (2023). Artificial intelligence to automate the systematic review of scientific literature. *Computing*, *105*(10), 2171-2194. https://doi.org/10.1007/s00607-023-01181-x

Korzynski, P., Mazurek, G., Krzypkowska, P., & Kurasinski, A. (2023). Artificial intelligence prompt engineering as a new digital competence: Analysis of generative AI technologies such as ChatGPT. *Entrepreneurial Business and Economics Review*, *11*(3), 25-37. https://doi.org/10.15678/EBER.2023.110302

Kwak, J., Jamal, A., Jones, B., Timmerman, G. M., Hughes, B., & Fry, L. (2022). An Interprofessional Approach to Advance Care Planning. *Am J Hosp Palliat Care*, *39*(3), 321-331. https://doi.org/10.1177/10499091211019316

Lee, D., & Palmer, E. (2025). Prompt engineering in higher education: a systematic review to help inform curricula. *International Journal of Educational Technology in Higher Education*, *22*(7). https://doi.org/10.1186/s41239-025-00503-7

Lee, J. M. G., Chan, C. Q. H., Low, W. C., Lee, K. H., and Low, L. L. (2020). Health-seeking behaviour of the elderly living alone in an urbanised low-income community in Singapore. *Singapore Med J*, *61*(5), 260-265. https://doi.org/10.11622/smedj.2019104

Leite, D. F. B., Padilha, M. A. S., & Cecatti, J. G. (2019). Approaching literature review for academic purposes: The Literature Review Checklist. *Clinics (Sao Paulo)*, *74*, e1403. https://doi.org/10.6061/clinics/2019/e1403

Levine, S., Beck, S. W., Mah, C., Phalen, L., & PIttman, J. (2024). How do students use ChatGPT as a writing support? *Journal of Adolescent & Adult Literacy*, *n/a*(n/a). https://doi.org/10.1002/jaal.1373

Li, Z. (2023). The dark side of chatgpt: Legal and ethical challenges from stochastic parrots and hallucination. *arXiv preprint arXiv:2304.14347*.

Macri, A., Spong, G. (2023). *Crafting prompt sandwiches for generative AI*. Retrieved 12 February from https://www.elastic.co/blog/crafting-prompt-sandwiches-generative-ai

Mali, Y.C.G. (2025). Exploring the use of ChatGPT in EFL/ESL writing classrooms: a systematic literature review. *Journal of Language and Education*, 11(2), 137-156. https://doi.org/10.17323/jle.2025.21793

Marzuki, Widiati, U., Rusdin, D., Darwin, & Indrawati, I. (2023). The impact of AI writing tools on the content and organization of students' writing: EFL teachers' perspective. *Cogent Education*, *10*(2), 2236469. https://doi.org/10.1080/2331186X.2023.2236469

Okaiyeto, S., & Xiao, H.-W. (2023). Generative AI in education: To embrace it or not？. *International Journal of Agricultural and Biological Engineering*, *16(3)*, 285-286. https://doi.org/10.25165/j.ijabe.20231603.8486

Ondrusek, A. L. (2012). What the Research Reveals about Graduate Students' Writing Skills: A Literature Review. *Journal of Education for Library and Information Science*, *53(3)*, 176-188.

Pearson, H. (2024). Can AI review the scientific literature - and figure out what it all means? *Nature*, *635*(8038), 276-278. https://doi.org/10.1038/d41586-024-03676-9

Raitskaya, L. & Tikhonova, E. (2025). Enhancing critical thinking skills in ChatGPT-human interaction: a scoping review. *Journal of Language and Education*, 11(2), 5-12. https://doi.org/10.17323/jle/2025.27387

Rewhorn, S. (2018). Writing your successful literature review. *Journal of Geography in Higher Education*, *42*(1), 143-147. https://doi.org/10.1080/03098265.2017.1337732

Sahoo, P., Singh, A., Saha, S., Jain, V., Mondal, S., & Chadha, A. (2024). *A Systematic Survey of Prompt Engineering in Large Language Models: Techniques and Applications*. https://doi.org/10.13140/RG.2.2.13032.65286

Tassoti, S. (2024). Assessment of Students Use of Generative Artificial Intelligence: Prompting Strategies and Prompt Engineering in Chemistry Education. *Journal of Chemical Education*, *101*(6), 2475-2482. https://doi.org/10.1021/acs.jchemed.4c00212

U.S. Department of Education, O. o. E. T., A. (2023). *Artificial Intelligence and the Future of Teaching and Learning: Insights and Recommendations*.

Wagner, G., Lukyanenko, R., & Paré, G. (2022). Artificial intelligence and the conduct of literature reviews. *Journal of Information Technology*, *37*(2), 209-226. https://doi.org/10.1177/02683962211048201

Walter, L., & Stouck, J. (2020). Writing the Literature Review: Graduate Student Experiences. *The Canadian Journal for the Scholarship of Teaching and Learning*, *11*(1). https://doi.org/10.5206/cjsotl-rcacea.2020.1.8295

Wang, M., Wang, M., Xu, X., Yang, L., Cai, D., & Yin, M. (2024). Unleashing ChatGPT's Power: A Case Study on Optimizing Information Retrieval in Flipped Classrooms via

PROMPT ENGINEERING TO IMPROVE LITERATURE REVIEWSPrompt Engineering. *IEEE Transactions on Learning Technologies*, *17*, 629-641. https://doi.org/10.1109/TLT.2023.3324714

White, J., Fu, Q., Hays, S., Sandborn, M., Olea, C., Gilbert, H., Elnashar, A., Spencer-Smith, J., & Schmidt, D. C. (2023). A prompt pattern catalog to enhance prompt engineering with chatgpt. *arXiv preprint arXiv:2302.11382*.

Winchester, C. L., & Salji, M. (2016). Writing a literature review. *Journal of Clinical Urology*, *9*(5), 308-312. https://doi.org/10.1177/2051415816650133

Yang, K., Raković, M., Liang, Z., Yan, L., Zeng, Z., Fan, Y., Gašević, D., & Chen, G. (2025). *Modifying AI, Enhancing Essays: How Active Engagement with Generative AI Boosts Writing Quality* Proceedings of the 15th International Learning Analytics and Knowledge Conference, https://doi.org/10.1145/3706468.3706544

Younas, A., & Ali, P. (2021). Five tips for developing useful literature summary tables for writing review articles. *Evid Based Nurs*, *24*(2), 32-34. https://doi.org/10.1136/ebnurs-2021-103417

Zala, K., Acharya, B., Mashru, M., Palaniappan, D., Gerogiannis, V., Kanavos, A., & Karamitsos, I. (2024). Transformative Automation: AI in Scientific Literature Reviews. *International Journal of Advanced Computer Science & Applications*, *15*(1) https://doi.org/10.14569/IJACSA.2024.01501122

Zentennial, F. (2024). *Unlocking the Power of COSTAR Prompt Engineering: A Guide and Example on converting goals into system of actionable items*. Retrieved 25 March from https://medium.com/@frugalzentennial/unlocking-the-power-of-costar-prompt-engineering-a-guide-and-example-on-converting-goals-into-dc5751ce9875